\documentclass[final,3p,times]{elsarticle}

\usepackage{amssymb}
\usepackage{amsmath}
\newcommand{\ket}[1]{\mathinner{| #1 \rangle}}
\newcommand{\e}{\mathrm e}
\newcommand{\im}{\mathrm i}

\usepackage{layout}

\journal{Physica A}

\begin{document}

\begin{frontmatter}



\title{Dynamical pairwise entanglement and two-point correlations in the three-ligand spin-star structure}


\author{M. Motamedifar}

\address{Department of Physics, Shahid Bahonar University of Kerman, Kerman, Iran}

\begin{abstract}
We consider the three-ligand spin-star structure through homogeneous Heisenberg interactions (XXX-3LSSS) in the framework of dynamical pairwise entanglement.
It is shown that the time evolution of the central qubit ``one-particle" state (COPS) results in the generation of quantum $W$ states at periodical time instants.
On the contrary, $W$ states cannot be generated from the time evolution of a ligand ``one-particle" state (LOPS). 
We also investigate the dynamical behavior of two-point quantum correlations as well as the expectation values of the different spin-components for each element in the XXX-3LSSS.
It is found that when a $W$ state is generated, 
the same value of concurrence among all qubits arises from $xx$ and $yy$ two-point quantum correlations. On the opposite, $zz$ quantum correlation between any two qubits vanishes at these time instants.
\end{abstract}


\end{frontmatter}


\section{Introduction}
\label{intro}

The development of the theory of quantum mechanics has helped us to understand the properties of low dimensional structures in details. Consequently, in the last decade, accessing, probing and engineering finite size chains of qubits have been practically considered in the literature~\cite{hirjibehedin2006spin, brune2006assembly}. In fact, the potential for scalability and miniaturization facilitates using spin qubits practically. For example, they can contribute to implementing quantum processors~\cite{shulman2012, joulain2016quantum}. Many research groups have considered finite size spin chains from quantum entanglement point of view~\cite{amse09exp, bar2010expe, monz201114}.

The theory of quantum entanglement is a central concept in a wide range of quantum information processing tasks and other different kinds of uniquely quantum mechanical phenomena. 
Quantum entanglement can be observed when the quantum state of a single part of a compound system cannot be described independent of the others.

In comparison with studies devoted to the static quantum entanglement, the dynamical point of view of entanglement has been less considered in the last two decades~\cite{wang2001enta, amico2004dynam, mintert05mea, hazzard14quan, qian2014pre, kuzmak15ge, wick2015ent}. As reported in Ref.~\cite{wang2001enta}, the quantum $W$ entangled states could be generated from time-evolving a ``one-particle" state in three- or four-qubit spin chains. A generalized form of an entangled quantum $W$ state for $N$ qubits~\cite{dur2000three, wang2001enta} can be given by 

\begin{equation}
\ket{W_N} = \frac{1}{\sqrt{N}} \sum_{j=1}^N \e^{\im \theta_j}
\ket{00 \ldots 010 \ldots 00}_j
\end{equation}
where $N$ is the number of particles and $\ket{00 \ldots 010 \ldots 00}_j$
represents a state with a ``1'' on the $j$-th position and zeros everywhere
else and $\theta_{i}$s ($i=1,..,N$) are some phases. As it shown by D$\ddot{u}$r \textit{et al}.~\cite{dur2000three} the concurrences~\cite{woot97, woot98, woot20} between any two qubits entangled by a quantum $W$ state are all equal to $2/N$. The entanglement of a $W$ state is not fragile in the sense of particle loss, i.~e., by tracing out one qubit, the state of remaining N-1 qubits can be again a maximally entangled state~\cite{guo2002scheme}.
There are some theoretical schemes and practical experiments for generating quantum $W$ states in the different approaches~\cite{guo2002scheme, yamamoto02, xiang2003scheme, eibl2004experimental, jin2009generation, zhang2014generation, hu2015w}. On the other hand, there have been several proposals for exploiting $W$ states in the quantum information processing~\cite{gorbachev2003can, joo2003quantum, gorbachev2005, cao2005, li2007supervised, li2007states}. 
Recently, an analyzer~\cite{zhu2015w} has been designed to distinguish between different $W$ states for a four-qubit system. They also mentioned some plans for $W$ state utilization as one of the most important resources in quantum key distribution.
The quantum key distribution is one of the most important field of quantum communication whose task is to create a private key between two remote legitimate users along an insecure quantum channel~\cite{bennett1984quantum, lo2014secure}. 

Depending on quantum correlations between different elements of a quantum system, there are various kinds of measure tools such as concurrence~\cite{woot97, woot98, woot20} and negativity~\cite{horodecki1996separability, peres1996separability, vidal2002computable} to quantify the entanglement . 
The concurrence has been widely used for measuring the pairwise entanglement between any two qubits in spin chains~\cite{li2006thermal, wang2007enta, an2007enta, wang2007fent, guo2009pair,  hide2012co, wahara2013, soltani2014q, fumani2016ma, mahmoudi2016, csahint2016en, gebrem2016dynam}. 

In addition to spin chains, which have been widely studied in one or quasi one-dimensional materials, there are spin-star structures (networks) which have attracted much interest recently~\cite{breuer2004non, palumbo2006quantum, krovi2007non, ferraro2008non, ross2008bang, wan2009tunable, arshed2010channel, JPHYSB10tripa, mili11genuine, Ma13tripartite, behzadi13the, turkpencce2016photonic}. These kinds of structures contain a central spin interacting with all outer spins which can be called ligands. On the other hand, ligands don't interact with each other directly. 
Such systems have shown a great potential in the realization of spin-qubit quantum computation because of their high symmetry~\cite{breuer2004non, krovi2007non, ferraro2008non}. Also due to their strong non-Markovian behavior, spin-star networks are an excellent models to study the decoherence of a single nitrogen-vacancy (NV) center coupled to a bath of nuclear spins in diamond~\cite{reinhard2012tuning, hanson2008coherent}. 
In addition, spin-star networks have been considered from quantum entanglement and quantum cloning point of view~\cite{hutton2002comparison, hutton2004mediated, de2004quantum, chen2010imp}. Moreover, the dynamical behavior of the entanglement between two non-interacting qubits in a bath of spin star structure has been considered in Ref.~\cite{palumbo2006quantum}.

\begin{figure}[b]
	\centering
	\includegraphics[width=0.30\linewidth]{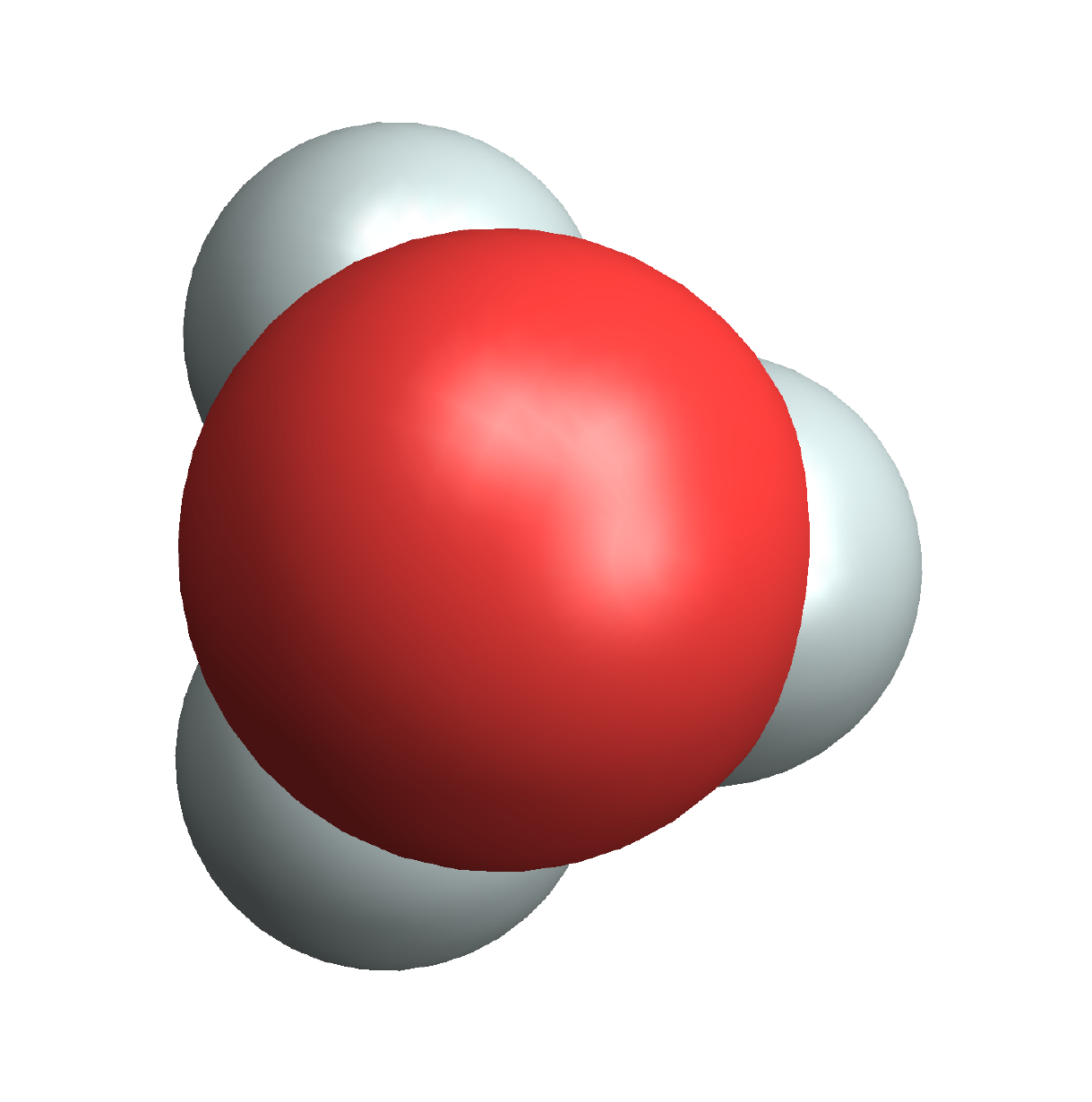}
	\caption{Schematic picture of a 3LSSS.}
	\label{fig:sfig11}
\end{figure}

One version of spin-star structures contains three ligands (3LSSS). The 3LSSS (Fig.~\ref{fig:sfig11}) through Heisenberg couplings or Dzyaloshinskii-Moryia (DM) interactions has been considered from thermal entanglement point of view. For instance, in Ref.~\cite{wan2009tunable}, pairwise entanglement between different parts of a 3LSSS with coupled microcavities in a thermal equilibrium has been studied. Ref.~\cite{JPHYSB10tripa} has paid attention to tripartite entanglement in a 3LSSS in which the central particle interacts with ligands through XY Heisenberg interactions. Moreover, tripartite entanglement has been focused again in a situation in which there is a DM interaction in the $z$-axis direction on each leg~\cite{Ma13tripartite}. A 3LSSS with interactions defined in Ref.~\cite{JPHYSB10tripa} has been considered again by authors of Ref.~\cite{behzadi13the} considering tripartite quantum discord. The pairwise thermal entanglement and quantum discord in an anisotropic 3LSSS have been investigated to detect the quantum phase transitions~\cite{motamedi2016}. 
However, the 3LSSS has not been so far investigated from a dynamical point of view of entanglement. 
In this case, some basic questions could be raised. For example, compared with a straight chain similar
to that was discussed in Ref.~\cite{wang2001enta}, what is the effect of this modified configuration on the dynamical behavior of the entanglement, when the structure has an additional Heisenberg interaction in the $z$-direction. 
  
The star network of spins with homogeneous Heisenberg interactions on each leg has many symmetric properties due to its invariance under the exchange of any two ligands. In fact, homogeneous Heisenberg model (XXX model) and anisotropic version of that (XXZ) are the heart of the integrable models of magnetism in low dimensional materials. 
In this work, we are interested to address the exact time evolution of one-particle states in the XXX-3LSSS which is described by the following Hamiltonian:
\begin{equation}
H=J\sum_{n=2}^N \overrightarrow{S}_{1}. \overrightarrow{S}_{n},  \label{eq:hamiltonixxx}
\end{equation}
where $\overrightarrow{S}_{1}$ is the spin vector of the spin-$\frac{1}{2}$ central qubit, and $\overrightarrow{S}_{n}$ $(n=2,3,4)$ shows the spin vector of the $\textit{n}$th spin-$\frac{1}{2}$ ligand, and here $N=4$. The parameter $J$ denotes the exchange coupling between the central qubit and ligands.
Investigation of the dynamical behavior of the pairwise entanglement reveals that $W$ states can be generated just due to the time evolution of the COPS. 
The outline of this work is as follows: 
In Sec.~\ref{sec:tecp} we present an explicit form of the time evolution of the COPS. 
In Sec.~\ref{sec:PWEpc}, we investigate the concurrences between any two qubits. Two-point quantum correlations are calculated in Sec.~\ref{sec:corfuncp}.
We also study the time evolution of an arbitrary LOPS in Sec~\ref{sec:ligands}. In Secs~\ref{sec:tel} and~\ref{sec:CFUl} the dynamical behavior of concurrences and correlation functions for the case of LOPS time-evolving are studied respectively. Conclusions and perspectives are presented in Sec.~\ref{sec:conclu}. 

\section{The time evolution of the COPS}
\label{sec:tecp}
Let us start working by the COPS applied upon by the time evolution operator which is given by $U(t)=e^{\frac{-iHt}{\hbar}}$. Where $H$ is the Hamiltonian (Eq.~\ref{eq:hamiltonixxx}) in which $J=1$. Without losing any generality and for simplicity, we consider $\hbar=1$ through this draft.
With the help of the raising operator ($S^{+}=S_x+iS_y$), a one-particle quantum state~\cite{wang2001enta, amico2004dynam} in a four-qubit system can be given by $S^{+}_{n}|0000\rangle$ where $(n=1,2,3,4)$.
Since one-particle states are not the eigenvectors of system's Hamiltonian, thus their time evolved format generally reads as Eq.~\ref{eq:tipsi}.
\begin{eqnarray}
|\psi(t) \rangle =b_{1}(t)|1000\rangle+b_{2}(t)|0100\rangle+b_{3}(t)|0010\rangle+b_{4}(t)|0001\rangle,
\label{eq:tipsi}
\end{eqnarray}
here $|\psi(t) \rangle$ is $U(t)S_{n}^{+}|0000\rangle$.
Whether the initial state is the COPS or a LOPS, the coefficients ($b_{i}(t)$s) in Eq.~\ref{eq:tipsi}, can be determined. In the COPS, the central spin is up and the ligand spins are down. The time evolution of the COPS can be given by
\begin{eqnarray}
|\psi(t)\rangle=
(\frac{1}{4} e^{-\frac{3 i t}{4}}+\frac{3}{4} e^{\frac{5 i t}{4}})|1000\rangle
+(\frac{1}{4} e^{-\frac{3 i t}{4}}-\frac{1}{4} e^{\frac{5 i t}{4}})\Big( |0100\rangle 
+|0010\rangle 
+|0001\rangle\Big) .
\label{eq:timeevolution1000}
\end{eqnarray}
As well as it can be seen in Eq.~\ref{eq:timeevolution1000}, the coefficients of various LOPSs are equal to each others and differ from that of the COPS.
\subsection{The pairwise entanglement after the COPS time evolution}
\label{sec:PWEpc}
The concurrence which has been suggested by Wootters~\textit{et al.}~\cite{woot97, woot98, woot20} is a quantifier of the pairwise entanglement between qubits $k$ and $j$ in a quantum system. Their proposed formula for the quantification of entanglement in a mixed state would be so promising. For a mixed quantum state, the density matrix can be written as:
\begin{equation}
\rho = \sum_{i} p_{i}|\phi_{i}\rangle\langle\phi_{i}|,
\end{equation}
where $p_{i}$, is the probability of the ket $|\phi_{i}\rangle$. For a pure state $p_{i}$ equals to $1$ and there is just one physical state to describe the system. 
In the case of tracing over two qubits $k$ and $j$ we can access to the reduced density matrix that is denoted by
\begin{equation}
\rho(k,j) = Tr_{k,j}(\rho),
\end{equation}
and then the concurrence between two qubits $\textit{k}$, and $\textit{j}$ is given by
\begin{equation}
C(k,j) = max\left\lbrace 2\lambda_{1}-\sum_{r}\lambda_{r}, 0\right\rbrace.
\label{eq:concu}
\end{equation}
In this equation, the $\lambda_{r}$s are square roots of eigenvalues of the matrix $R=\rho(k,j)\sigma_{y}\otimes \sigma_{y}~\rho^{*}(k,j)~\sigma_{y}\otimes\sigma_{y}$, in which $\sigma_{y}$ denotes the Pauli-y-matrix  and $\rho^{*}$ is the complex conjugate of $\rho$.

$0<C(k,j)<1$ implies that two qubits $\textit{k}$, and $\textit{j}$ are partially entangled. The $C(k,j)=1$ corresponds to a maximally entangled state as well as $C(k,j)=0$ implies to a completely disentangled state. For one-particle states, concurrence can be given~\cite{wang2001enta, amico2004dynam} by: 
\begin{eqnarray}
C(k,j)(t)=2|b_{k}(t)b_{j}(t)|,
\label{concu}
\end{eqnarray} 
where, $b_{i}(t)$s are the coefficients in Eq.~\ref{eq:tipsi}. 

In Fig.~\ref{fig:plot1000}, the dynamical behavior of the concurrence between two qubits $\textit{k}$ and $\textit{j}$ is exhibited by $C_{k-j}$. As shown in Fig.~\ref{fig:gerrucp} there are two kinds of pairs. One kind includes the central particle and a ligand, thus concurrence for this pair is shown by $C_{ucp-l}$. Another type of pairs consists two ligands. The concurrence for this type is shown by $C_{l-l}$. 
Concerning Eqs~\ref{eq:timeevolution1000} and~\ref{concu}, we can write the following equations for the time dependent functions of concurrences:
\begin{figure}[t]
	\centering
	\begin{minipage}[b]{0.38\textwidth}
		\includegraphics[width=\textwidth]{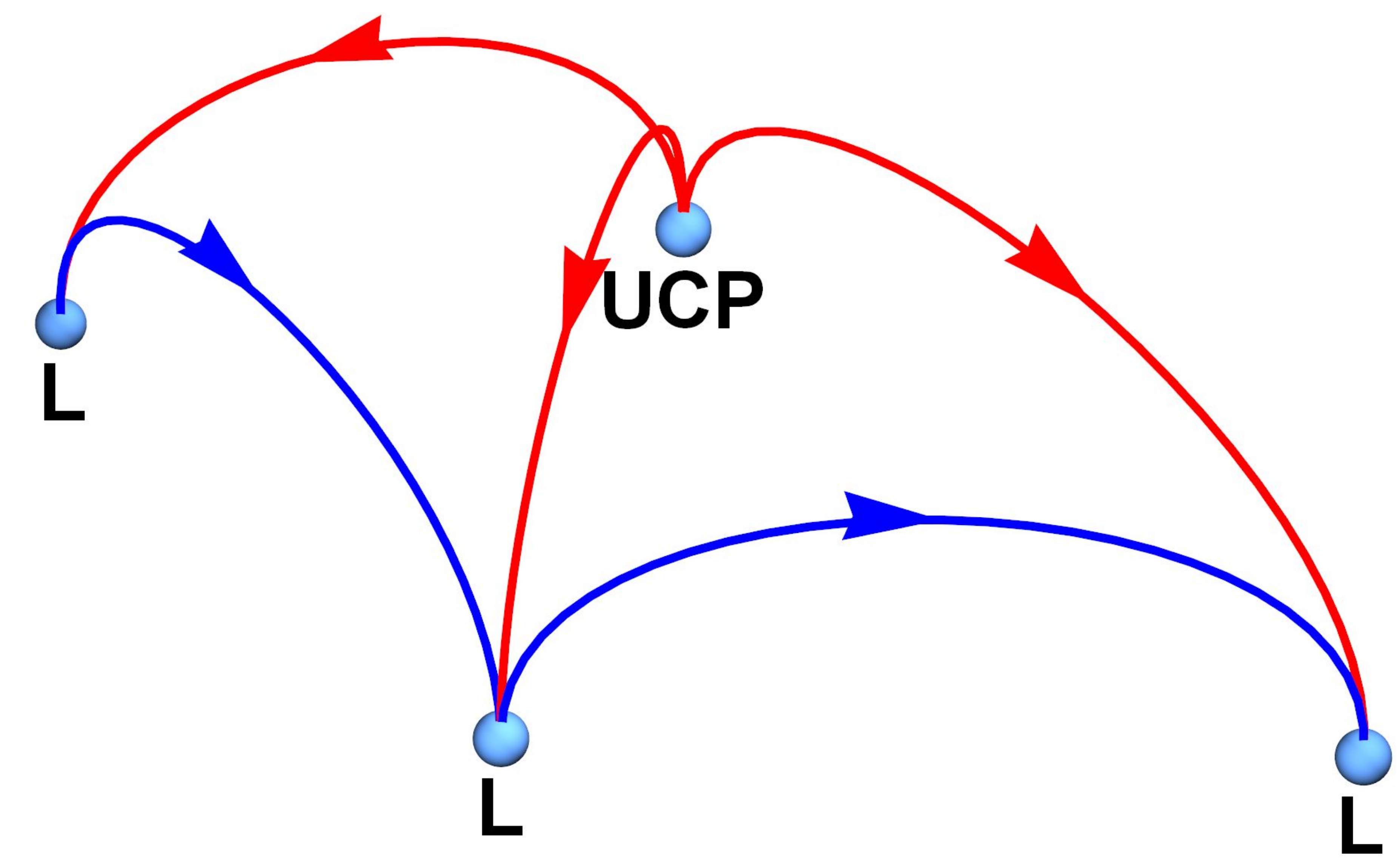}
		\caption{Schematic picture of different pairs in XXX-3LSSS in the situation of COPS time evolution. The central qubit whose one-particle state has been aplied upoun by U(t) is shown by UCP and any ligand presented by L.}
		\label{fig:gerrucp}
	\end{minipage}
	\hfill
	\begin{minipage}[b]{0.5\textwidth}
		\includegraphics[width=\textwidth]{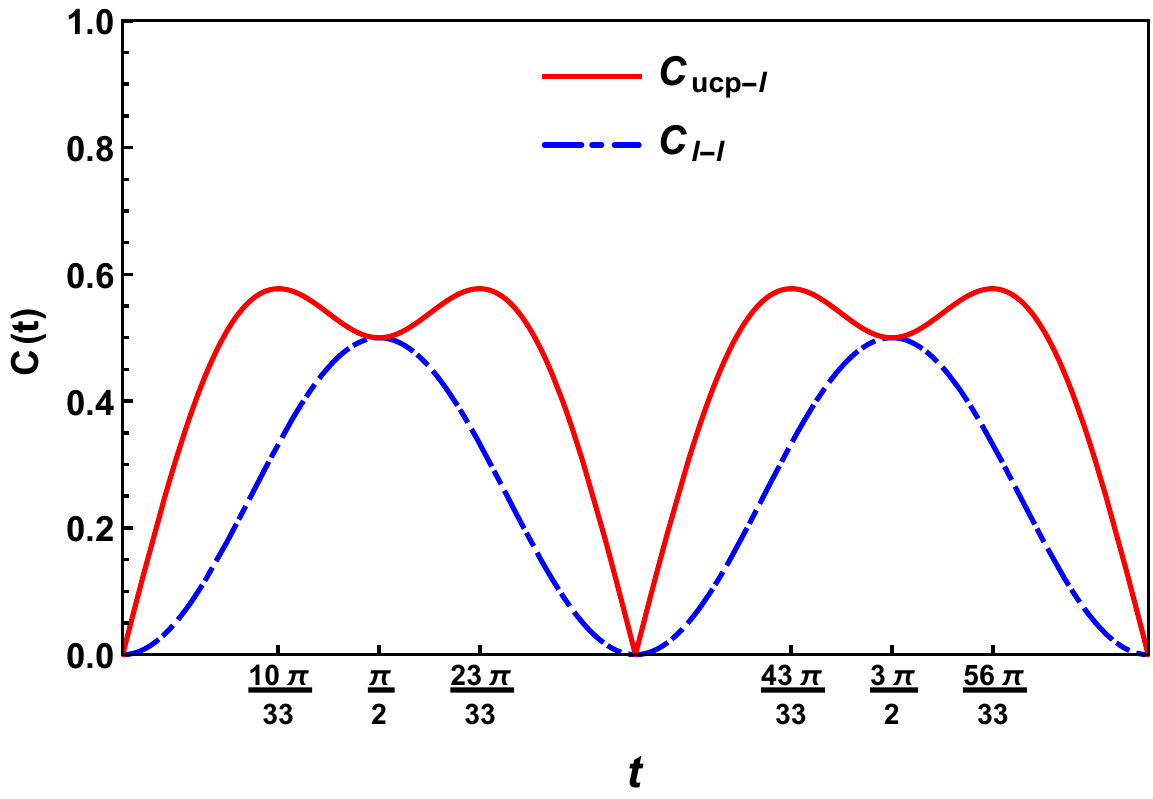}
		\caption{The time-dependence concurrences for various pairs.}
		\label{fig:plot1000}
	\end{minipage}
\end{figure}
\begin{eqnarray}
C_{ucp-l} & = & \frac{1}{2} \sqrt{\left(\sin ^4(t)+\sin ^2(2 t)\right)}\nonumber\\
C_{l-l} & = & \frac{\sin ^2(t)}{2}.
\label{ucpconco}
\end{eqnarray}
As Fig.~\ref{fig:plot1000} shows, concurrences alternate with a period of $\pi$. As depicted by the solid line in Fig.~\ref{fig:plot1000}, $C_{ucp-l}$ has peaks at certain time instants which can be given by Eq.~\ref{maxPWEucp-L}. It means that the central qubit maximally entangles with any ligand at these points. Here and following, throughout this work, $\textit{n}$ is an integer number greater than zero ($\textit{n}>0$). 
\begin{eqnarray}
t_{n}\simeq\frac{\pi  (2 n-1)}{4} +\frac{7\pi(-1)^{n+1}}{132}.
\label{maxPWEucp-L}
\end{eqnarray}
Also as shown with a dotted line in Fig.~\ref{fig:plot1000}, any two ligands maximally entangle with each other at certain time instants which can be given by Eq.~\ref{eq:maxll}.
\begin{equation}
t_{n}=\frac{\pi(2n-1)}{2}.
\label{eq:maxll}
\end{equation}
On the other hand, exactly at these points, $C_{ucp-l}$ equals to $C_{l-l}$ with a value of $0.5$ which implies that quantum $W$ states can be generated. Henceforth, these time instants are abbreviated as TWS. In the other words, at TWS the quantum $W$ states can be explicitly written as the following equation:
\begin{eqnarray}
|\psi(t_{n}=\frac{\pi(2n-1)}{2})\rangle=
\frac{1}{2} (-1)^{\frac{4+n}{8}}\Big(|1000\rangle
-|0100\rangle 
-|0010\rangle 
-|0001\rangle\Big).
\label{eq:t1000npi2}
\end{eqnarray}
For $n=1$, ($t_{1}=\frac{\pi}{2}$), the quantum state can be written as Eq.~\ref{eq:ti1000pi2} where $(-1)^{\frac{5}{8}}=i\ cos(\frac{\pi}{8})-\ sin(\frac{\pi}{8})$.
\begin{eqnarray}
|\psi(t=\frac{\pi}{2})\rangle=
\frac{1}{2} (-1)^{\frac{5}{8}}\Big(|1000\rangle
-|0100\rangle
-|0010\rangle 
-|0001\rangle\Big),
\label{eq:ti1000pi2}
\end{eqnarray} 

In order to consider the probability of the presence of each one-particle state at time $t$, we must calculate the following function:
\begin{equation}
P_{i}(t)=|\langle{i}|\psi(t)\rangle|^{2},
\label{eq:what}
\end{equation}
where $|i\rangle$ is the COPS or a LOPS, and $|\psi(t)\rangle$ is given by Eq.~\ref{eq:timeevolution1000}. The results of the above equation are written in Eqs.~\ref{eq:probabbility1000t} as probabilities for the different one-particle states.
\begin{figure}[t]
	\centering
	\begin{minipage}{0.49\textwidth}
		\includegraphics[width=\textwidth]{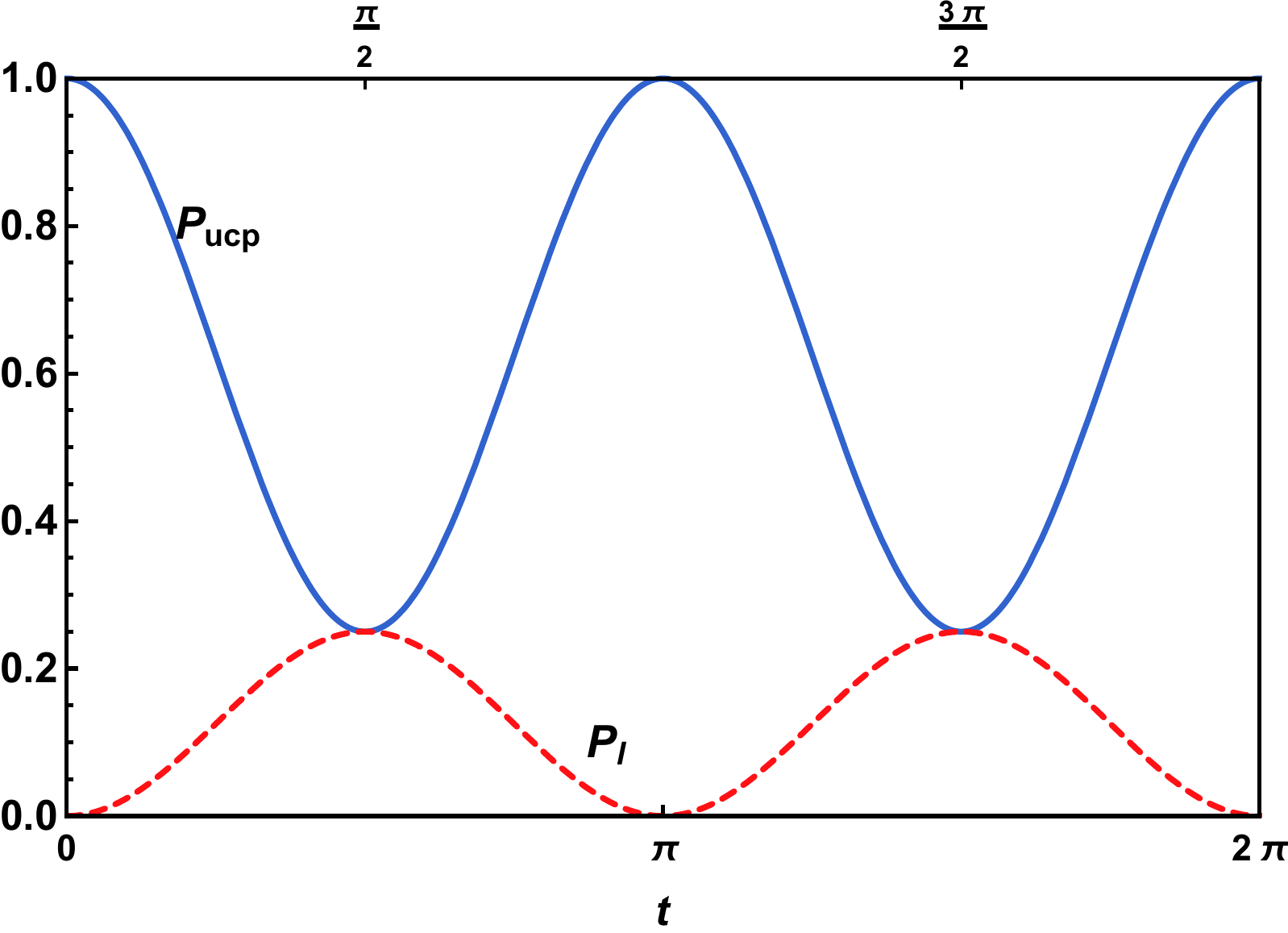}
		\caption{Probabilities of different one-particle states of XXX-3LSSS.}
		\label{fig:prob1000t}
	\end{minipage}
	\hfill
	\begin{minipage}{0.49\textwidth}
		\includegraphics[width=\textwidth]{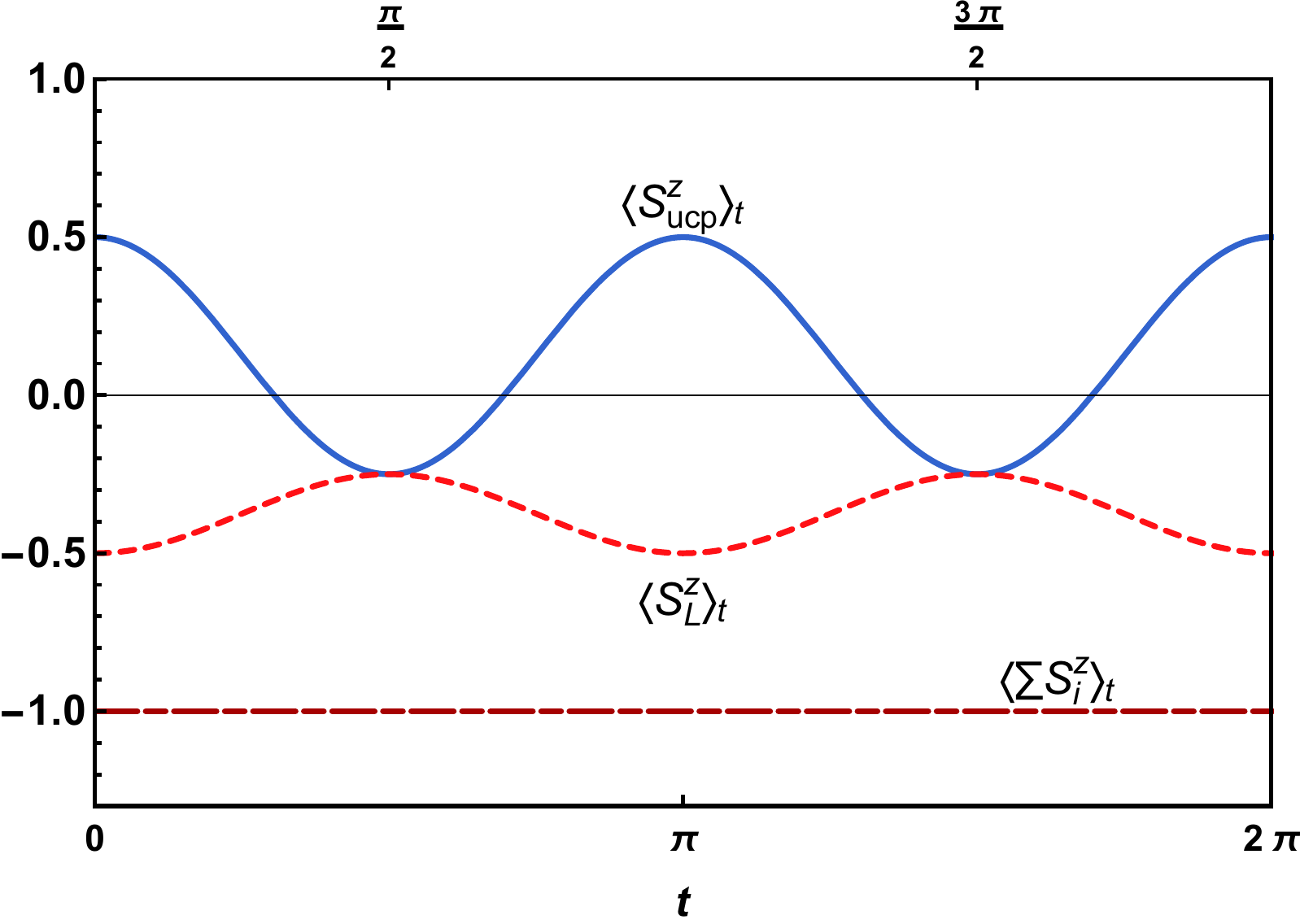}
		\caption{Expectation value of $z$-spin component of different qubits.}
		\label{fig:chashmeszi1000}
	\end{minipage}
\end{figure}
\begin{eqnarray}
P_{ucp}(t)&=&\frac{1}{8} (3 \cos (2 t)+5)\nonumber\\
P_{l}(t)&=&\frac{\sin ^2(t)}{4}.
\label{eq:probabbility1000t}
\end{eqnarray}
Here $P_{ucp}(t)$ is the probability of the COPS and $P_{l}(t)$ is the probability of any LOPS which are plotted in Fig.~\ref{fig:prob1000t}. It is clear to see that at $t=n\pi$ just the COPS is present and any LOPS is absent. Moreover we can see that at TWS, the probabilities of the COPS and any LOPS are equal to $0.25$. 
\subsection{Two-point quantum correlations after the COPS time evolution}
\label{sec:corfuncp}
To gain a more profound understanding about the past outcomes, let us start calculating the expectation values of the $x$- and $z$-spin components of the different qubits. Exploiting Eq.~\ref{eq:timeevolution1000}, the time evolution of $\langle S^{\alpha}_{i}\rangle_{t}$ can be given as:
\begin{eqnarray}
\langle S^{z}_{ucp}\rangle_{t}= \frac{1}{8} (3 \cos (2 t)+1)\notag\\
\langle S^{z}_{l}\rangle_{t}=- \frac{1}{8} (\cos (2 t)+3),
\label{szt1000}
\end{eqnarray}
which are presented in Fig.~\ref{fig:chashmeszi1000}.
Because $\langle S_{i}^x \rangle_t$ and $\langle S_{i}^y \rangle_t$ are equal to zero, so they are not shown in this plot. 

Moreover, focusing on Eqs.\ref{eq:probabbility1000t} and \ref{szt1000}, it is found that $\langle S^{z}_{i}\rangle_{t}=P_{i}(t)-0.5 $ ($i$ can be $ucp$ and $l$ which indicates respectively the central qubit and one of three ligands). In addition, since $\sum_{i}P_{i}(t)=1$, thus we can conclude that the total magnetization, $\sum_{i=1}^{4}\langle S^{z}_{i}\rangle_{t}$, equals to $-1$ and is constant with time as it is plotted in Fig.~\ref{fig:chashmeszi1000}. Furthermore, as it is clear in Fig.~\ref{fig:chashmeszi1000}, the $\langle S^{z}_{i}\rangle_{t}$s are all equal to each other at TWS.

To complement our studies about the COPS time evolution we consider the two-point correlation function between various spin components of different qubits ($W^{\alpha\alpha}_{\imath\jmath}$) which can be defined by the following equation:
\begin{eqnarray}
W^{\alpha\alpha}_{\imath\jmath}(t)=\langle\psi(t)|S_{\imath}^{\alpha}S_{\jmath}^{\alpha}|\psi(t)\rangle,
\label{correlationW}
\end{eqnarray} 
where $\alpha$ can be $x$ or $z$, and $|\psi(t)\rangle$ is given by Eq.~\ref{eq:timeevolution1000}. The results of calculating correlation functions are written in Eqs.~\ref{eq:corzxucp} for any two possible pairs of the XXX-3LSSS. Since the correlation for $y$ axis has a similar behavior of correlation in the $x$-direction, we just calculate the correlations in the $x$- and $z$-direction.
\begin{figure}[t]
	\centering
	\includegraphics[width=0.92\linewidth]{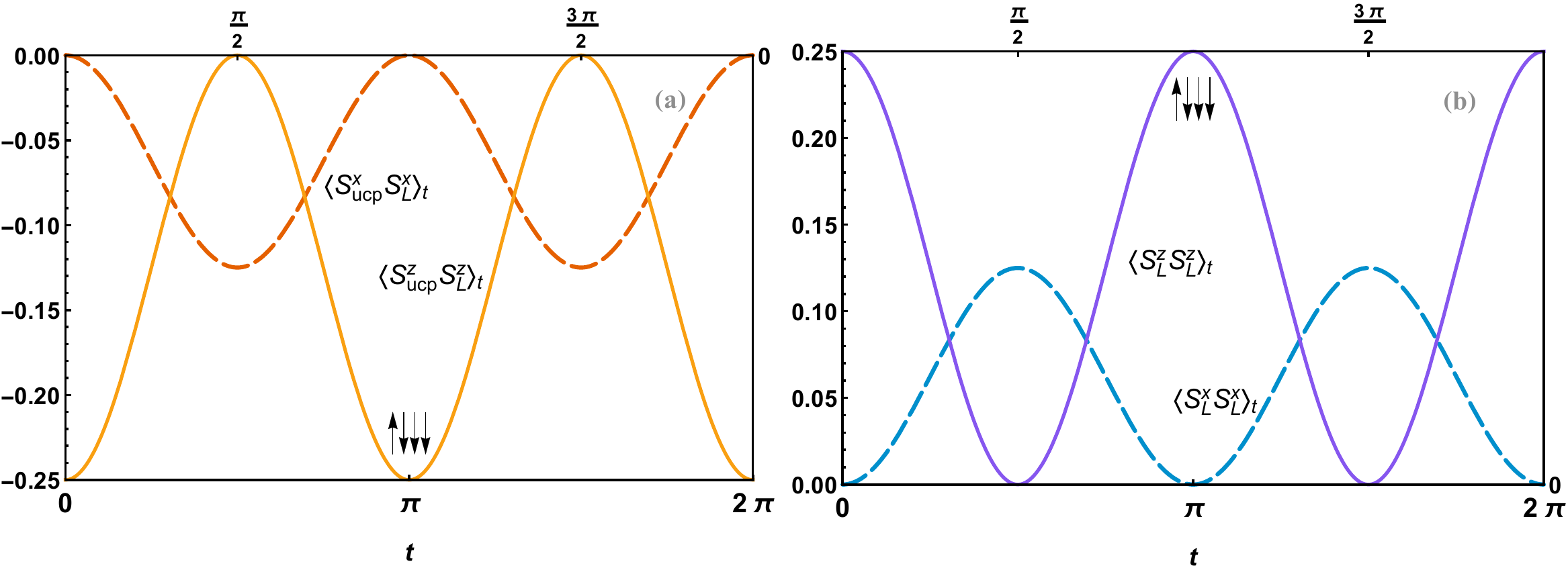}
	\caption{Two-point correlation functions of $z$- and $x$-spin components (a) between the UCP and a ligand, (b) between any two ligands.}
	\label{fig:corucp}
\end{figure}

\begin{eqnarray}
\langle S^{z}_{ucp}S^{z}_{L}\rangle_{t}&=&-\frac{1}{4} \cos ^2(t) \notag\\
\langle S^{x}_{ucp}S^{x}_{L}\rangle_{t}&=&-\frac{1}{8} \sin ^2(t)\notag\\
\langle S^{z}_{L}S^{z}_{L}\rangle_{t}&=&\frac{\cos ^2(t)}{4}\notag\\
\langle S^{x}_{L}S^{x}_{L}\rangle_{t}&=&\frac{\sin ^2(t)}{8}.
\label{eq:corzxucp}
\end{eqnarray}
In the above equations, $\langle S^{\alpha}_{ucp}S^{\alpha}_{L}\rangle_{t}$ stands for the correlation between $\alpha$th-spin component of UCP and a ligand, and $\langle S^{\alpha}_{L}S^{\alpha}_{L}\rangle_{t}$ is related to any two ligands. These quantities are plotted in Figs.~\ref{fig:corucp}. At $t=0$ and $t=n\pi$, the $z$-spin component of any two qubits are completely correlated ($\langle S^{z}_{ucp}S^{z}_{L}\rangle=|\langle S^{z}_{L}S^{z}_{L}\rangle|=0.25$). On the other hand, any two qubits are entirely disentangled at these time instants (Fig.~\ref{fig:plot1000}). 
As a matter of fact it can be said that at TWS, the concurrence between any two qubits is the result of $xx$ and $yy$ quantum correlations. On the contrary, $zz$ quantum correlation between any two qubits vanishes at TWS.
\begin{figure}[b]
	\centering
	\includegraphics[width=0.38\linewidth]{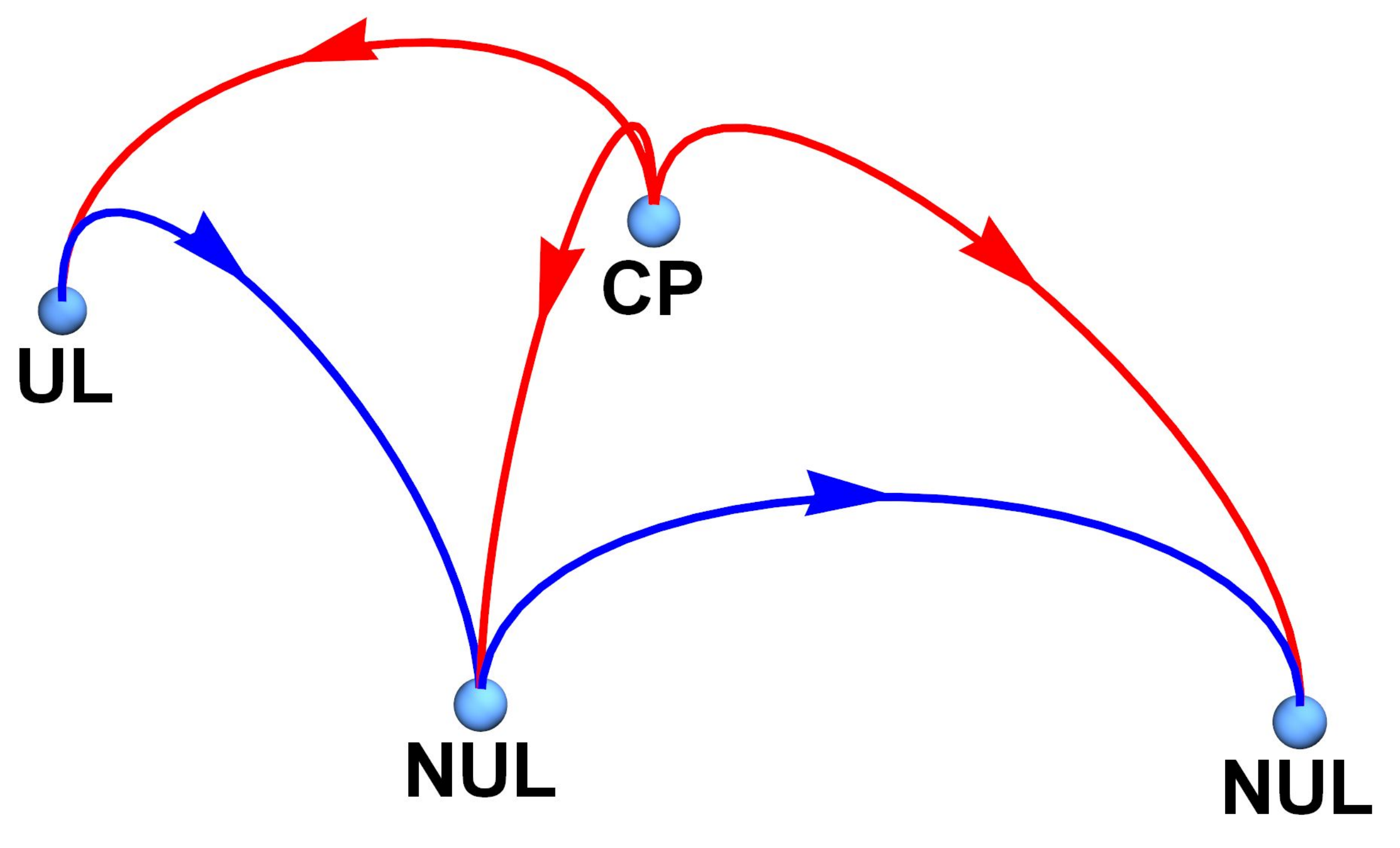}
	\caption{Schematic picture of the different pairs in the XXX-3LSSS for a LOPS time evolution. The central qubit is shown by CP. A ligand whose one-particle state has been acted upon by the time evolution operator is presented by UL and other ligands are exhibited by NUL.}
	\label{fig:gerr}
\end{figure}
\section{The time evolution of a LOPS} 
\label{sec:ligands}
We now move to study a LOPS time evolution. For our calculation, we pick $|0001\rangle$ as an arbitrary LOPS which will be acted upon by U(t). After calculation, $|\psi(t) \rangle=U(t)|0001\rangle$ can be written as:
\begin{eqnarray}
|\psi(t) \rangle&=&
(\frac{1}{4} e^{-\frac{3 i t}{4}}-\frac{1}{4} e^{\frac{5 i t}{4}})|1000\rangle
+(-\frac{1}{3} e^{-\frac{i t}{4}}+\frac{1}{4} e^{-\frac{3 i t}{4}}+\frac{1}{12} e^{\frac{5 i t}{4}})\Big(|0100\rangle 
+|0010\rangle\Big)\nonumber\\
&+&(\frac{2}{3} e^{-\frac{i t}{4}}+\frac{1}{4} e^{-\frac{3 i t}{4}}+\frac{1}{12} e^{\frac{5 i t}{4}})|0001\rangle
\label{eq:timeevolution0001}.
\end{eqnarray}
By considering the symmetric Hamiltonian (Eq.~\ref{eq:hamiltonixxx}), substituting other LOPSs instead of $|0001\rangle$ as the initial state, just exchanges coefficients in the above equation.    
\subsection{The pairwise entanglement after a LOPS time evolution} 
\label{sec:tel}

\begin{figure}[t]
	\centering
	\includegraphics[width=0.92\linewidth]{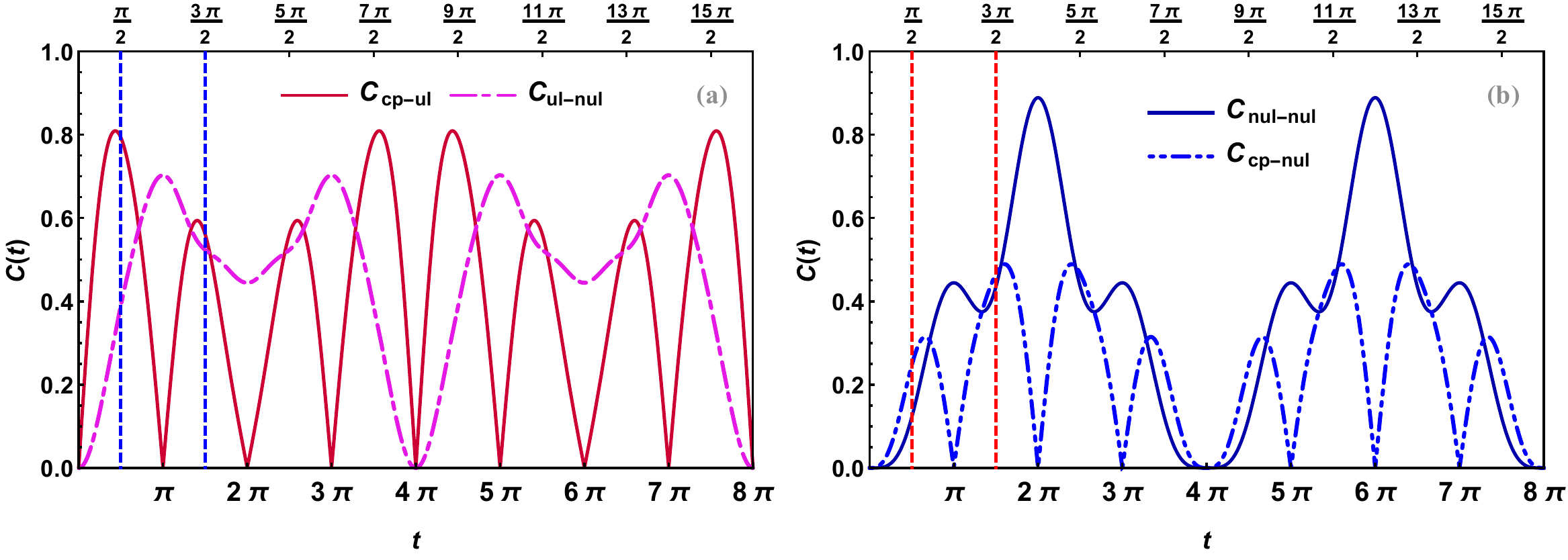}
	\caption{Concurrences between (a) UL and other qubits, (b) between two NULs or between a NUL and CP.}
	\label{fig:tUL-a-b}
\end{figure}

Focusing on the LOPS time evolution, we investigate the behavior of the concurrence between the different pairs which are schematically presented in Fig.~\ref{fig:gerr}.

The concurrences between the UL and other qubits are plotted in Fig.~\ref{fig:tUL-a-b}~(a). In addition, Fig.~\ref{fig:tUL-a-b}~(b) exhibits the concurrences between a NUL and other qubits.
As it can be seen, $C_{cp-ul}$ and $C_{cp-nul}$ are maximal at four instances ($t_{n}\simeq(2n-1)\pi/2$) over the course of one period, however they are not synchronous. On the opposite, we can see that the CP disentangles synchronously from all ligands at $t=n\pi$. Moreover, it is clear that, when a NUL maximally entangles with another NUL or the UL, it completely disentangles from the CP. 

\begin{figure}[t]
	\centering
	\includegraphics[width=0.92\linewidth]{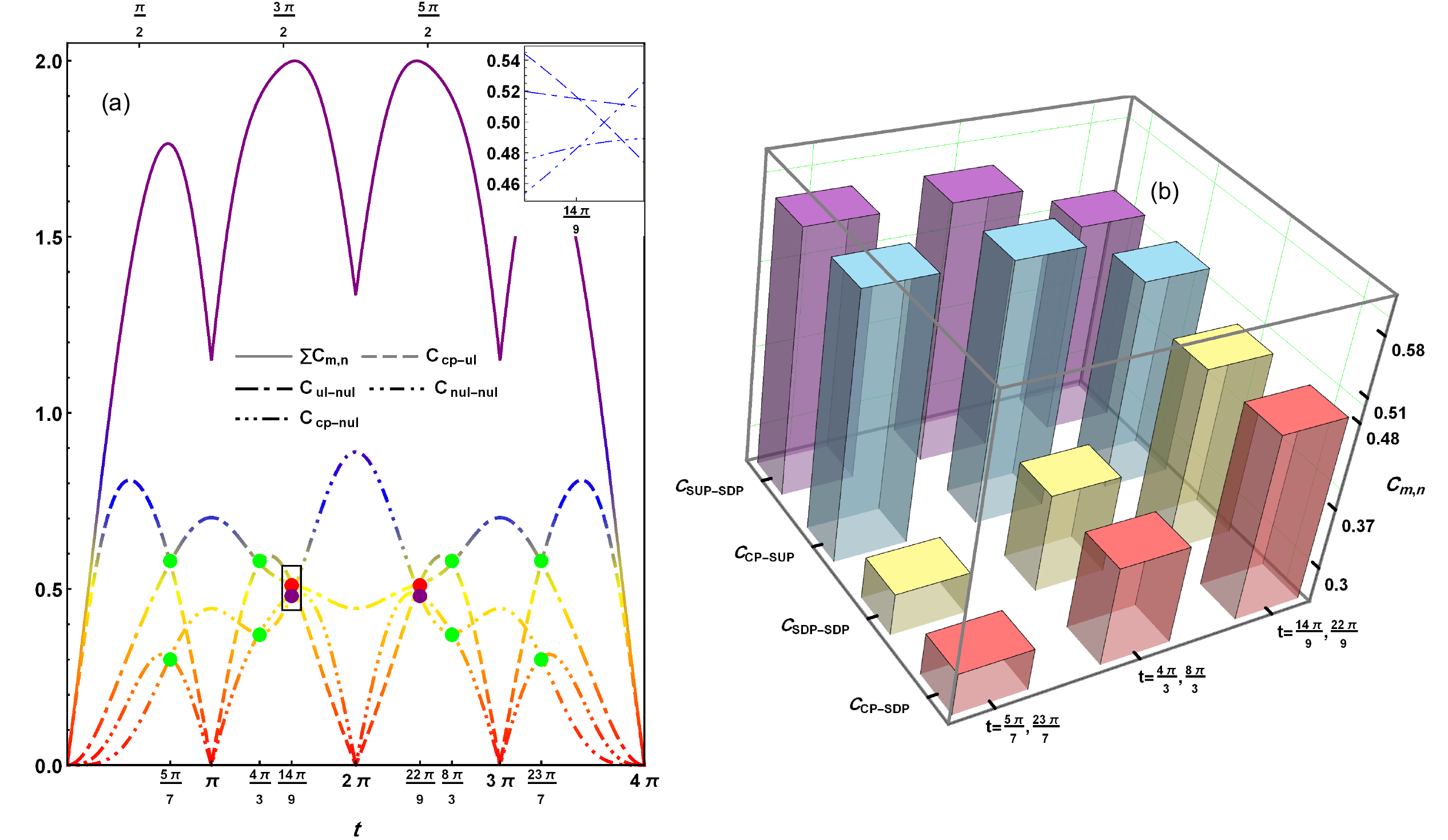}
	\caption{(a)~Concurrences for each pair, (b)~Concurrence levels of intersection points.}
	\label{fig:taghato}
\end{figure}

By looking at Figs.~\ref{fig:tUL-a-b} (a) and (b), it can be noticed that the concurrences cross with each other at some certain time instants. For this reason a question which can be asked here is: Can the quantum $W$ states be generated at those time instants?
In order to answer this question, we should check the equalization of concurrences for all pairs.  
Fig.~\ref{fig:taghato}~(a) which is plotted in one period, exposes some time instants at which different concurrences cross each other. Except for $t=14\pi/9$ and $t=22\pi/9$, at which intersections are very close to each other, all intersections are far away. For any pairs, the concurrence values on the location of the intersections are presented in Fig.~\ref{fig:taghato}~(b).
As both of Figs.~\ref{fig:taghato}~(a) and~(b) show, at $t=\left\lbrace 5\pi/7,  23\pi/7\right\rbrace $, $C_{cp-nul}$ and $C_{nul-nul}$ are in the same value of the concurrence ($C=0.30$), however, at these times, accompaniment of $C_{cp-ul}$ and $C_{ul-nul}$ comes to pass at the level of $C=0.58$.
Additionally, at $t=\left\lbrace 4\pi/3,  8\pi/3\right\rbrace $, $C_{cp-ul}$ and $C_{ul-nul}$ hold over the concurrence value at $C=0.58$, while $C_{cp-nul}$ and $C_{nul-nul}$ adjacently arise to $C=0.37$. 
Ultimately, as it is clear that at $t= \left\lbrace 14\pi/9, 22\pi/9\right\rbrace $, $C_{cp-ul}$ equals to $C_{ul-nul}$ with the value of $C=0.51$. On the other hand, $C_{cp-nul}$ and $C_{nul-nul}$ are equal to $C=0.48$ at these time instants. It means that however the concurrences are very close to each others but they are not exactly intersect. As it seen in the inset of Fig.~\ref{fig:taghato}~(a), although all concurrences are very close to each other at $t=14\pi/9$ but all four plots don't cross at the same level. These time instants at which concurrences are very close to each other but don't meet exactly at the same level, can be called as \textit{pseudo-$W$-state} time instants (PSTWS) and can be written as Eq.~\ref{nwst}. 
\begin{eqnarray}
t^{**}_{n}=\pi( 2 n-\frac{(9+5(-1)^{n})}{9}).
\label{nwst}
\end{eqnarray}
Moreover, the solid line in Fig.~\ref{fig:taghato}~(a) shows the summation of concurrences over all pairs on a course of one period. The highest amplitudes of the summation of concurrences takes place at PSTWS. Also, local minima of this function determine the time instances of disentangling the CP from all ligands.  

\begin{figure}[t]
	\centering
	\begin{minipage}[t]{0.49\textwidth}
		\includegraphics[width=\textwidth]{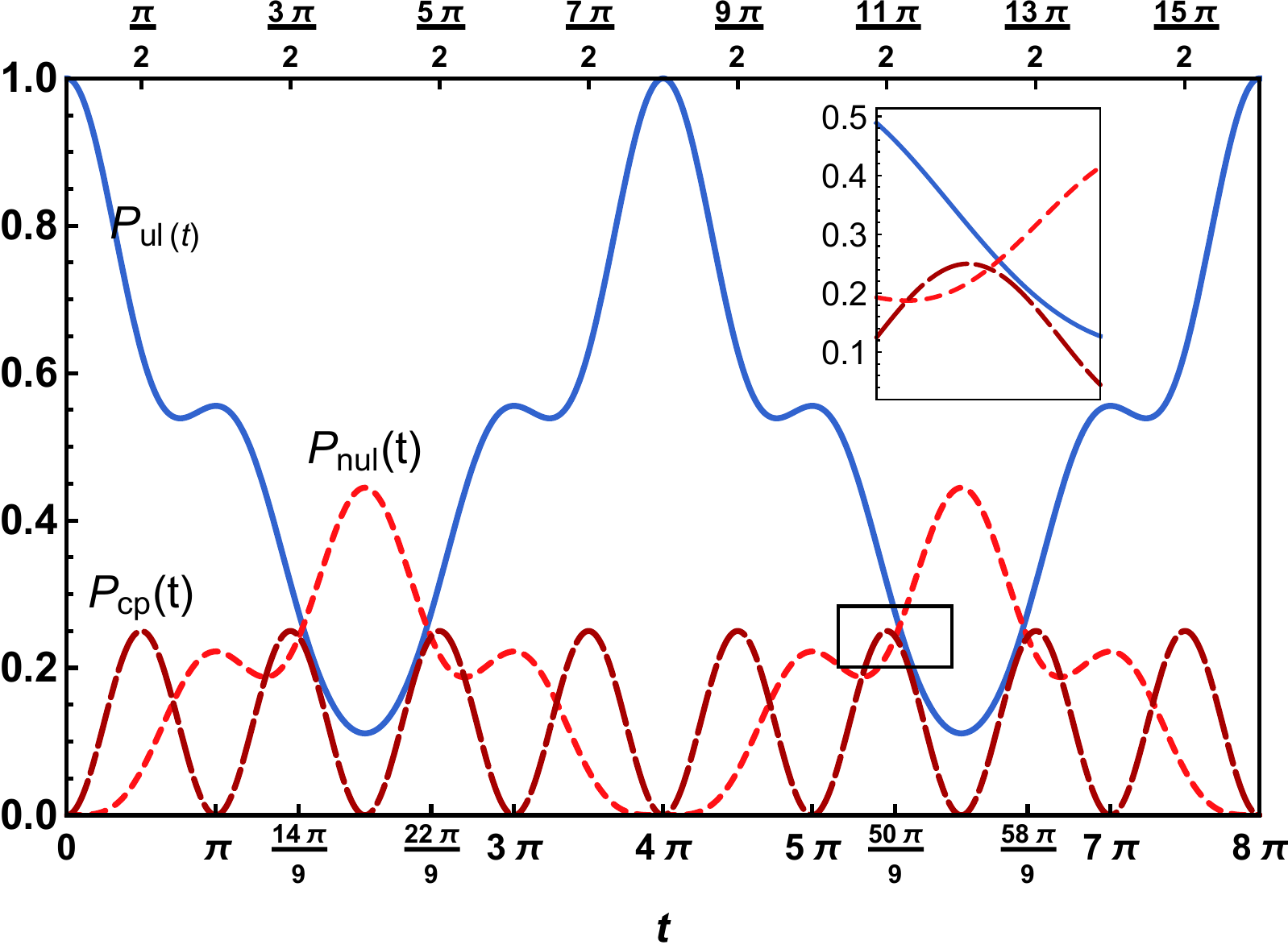}
		\caption{$P_{i}$s are the probabilities of one-paticle states in the situation of a LOPS time evolution ($i$ can be set to $cp$, $ul$ and $nul$).}
		\label{fig:probabbility0001t}
	\end{minipage}
	\hfill
	\begin{minipage}[t]{0.49\textwidth}
		\includegraphics[width=\textwidth]{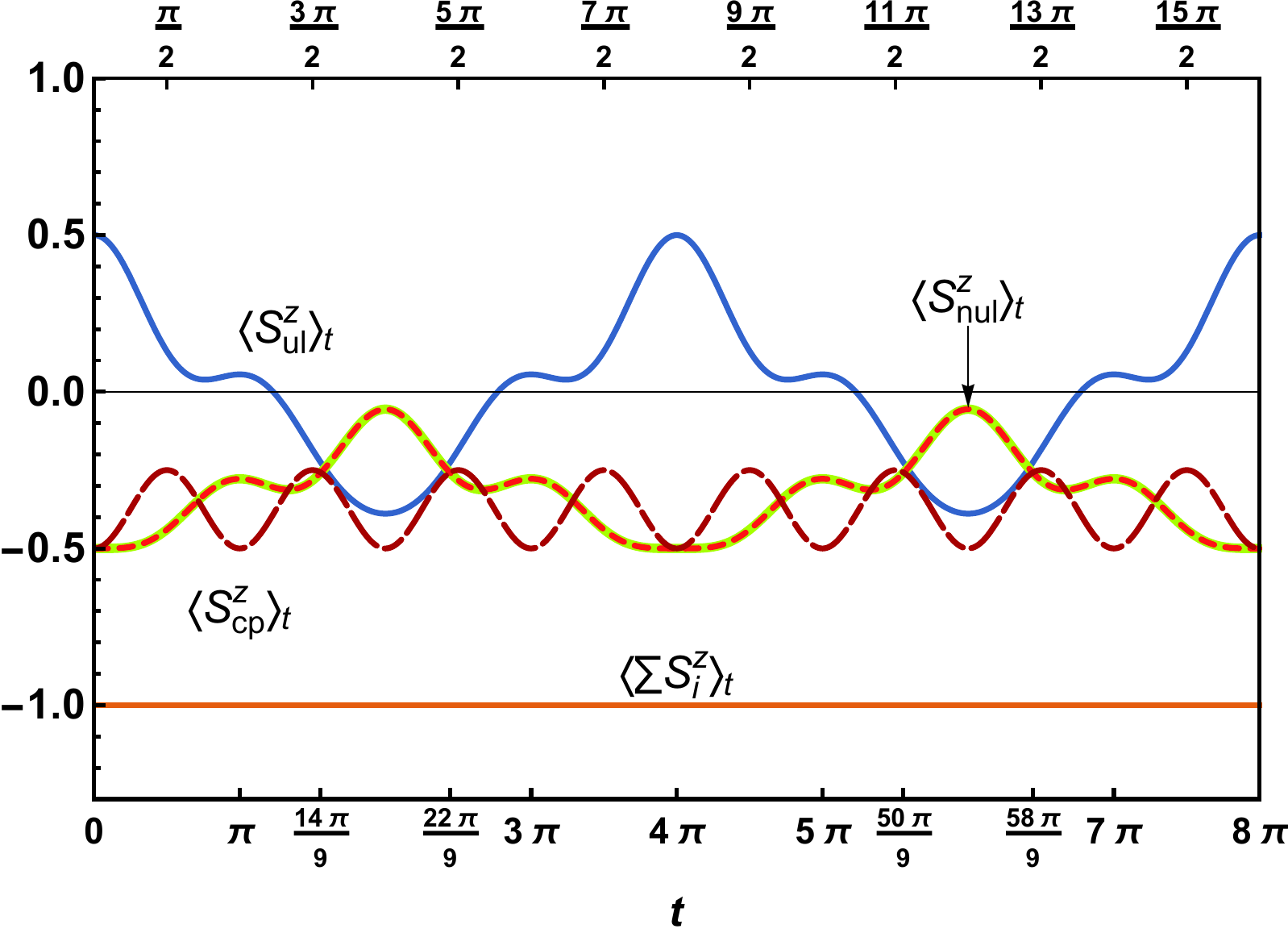}
		\caption{Expectation values of $S^{z}_{i}$ in the situation of a LOPS time evolution ($i$ can be set to $cp$, $ul$ and $nul$).}
		\label{fig:pshowchashmeszi}
	\end{minipage}
\end{figure}

In order to gain a deeper insight into the previous results, we now derive the probabilities of Eq.~\ref{eq:what} in which we insert the $|\psi(t)\rangle$ from Eq.~\ref{eq:timeevolution0001} and $i$ can be a one-particle state. Whether a one-particle state is the COPS or a LOPS the probabilities are given by the following equations:
\begin{eqnarray}
P_{cp}(t)&=&\frac{\sin ^2(t)}{4}\nonumber\\
P_{ul}(t)&=&\frac{1}{72} \left[{24} \cos \left(\frac{t}{2}\right)+8 \cos \left(\frac{3 t}{2}\right)+3 \cos (2 t)+37\right]\notag\\
P_{nul}(t)&=&\frac{2}{9} \sin ^4\left(\frac{t}{4}\right) \left[8 \cos \left(\frac{t}{2}\right)+3 \cos (t)+7\right].
\label{eq:propabilities}
\end{eqnarray} 
The probabilities are plotted in Fig.~\ref{fig:probabbility0001t}. As it is shown in this plot (and is clear in its inset), at $\tau^{**}$, the probabilities of 
any one-particle quantum states closely meet each others, however they are not completely equal.
\subsection{Two-point quantum correlation functions after a LOPS time evolution}
\label{sec:CFUl}
In this section, we discuss two-point correlations and the expectation value of the $z$ spin component in the case of a LOPS time evolution. 

Making use of Eq.~\ref{eq:timeevolution0001}, we can write Eqs.~\ref{sz0001t} for $\langle S^{z}_i\rangle_{t}$s which are presented in Fig.~\ref{fig:pshowchashmeszi}.
\begin{eqnarray}
\langle S^{z}_{cp}\rangle_{t}&=& -\frac{1}{8} (\cos (2 t)+3)\notag\\
\langle S^{z}_{ul}\rangle_{t}&=& \frac{1}{72}\Big[{24}\cos \left(\frac{t}{2}\right)+8 \cos \left(\frac{3 t}{2}\right)+3\cos (2 t)+1\Big]\notag\\
\langle S^{z}_{nul}\rangle_{t}&=&\frac{1}{72}\Big[{-12}\cos \left(\frac{t}{2}\right)-4 \cos \left(\frac{3 t}{2}\right)
+3\cos (2 t)-23\Big].
\label{sz0001t}
\end{eqnarray}
In addition, it is worth to consider that like the previous case (the COPS time evolution) probabilities are related to the expectation values by $\langle S_{i}^{z}\rangle_{t}=P_{i}(t)-0.5$ ($i$ can be $cp$, $l$ and $nul$). This fact then immediately results in $\sum_{i}\langle S_{i}^{z}\rangle_{t}=-1$. We can also see that $\langle S_{i}^{z}\rangle_{t}$s are very close to each other at PSTWS.
It is worth to mention that at certain times, the sign of $\langle S^{z}_{ul}\rangle_{t}$ changes from positive values to negative ones or vise versa. It is realized that the sign of $\langle S_{i}^{z}\rangle_{t}$ changes just for the qubit whose one-particle state applied upon by U(t). This situation happened to $\langle S^{z}_{ucp}\rangle_{t}$ (Sec.~\ref{sec:corfuncp}) as it was shown in Fig.~\ref{fig:pshowchashmeszi}.

\begin{figure}[t!]
	\centering
	\begin{minipage}{0.49\textwidth}
		\includegraphics[width=\textwidth]{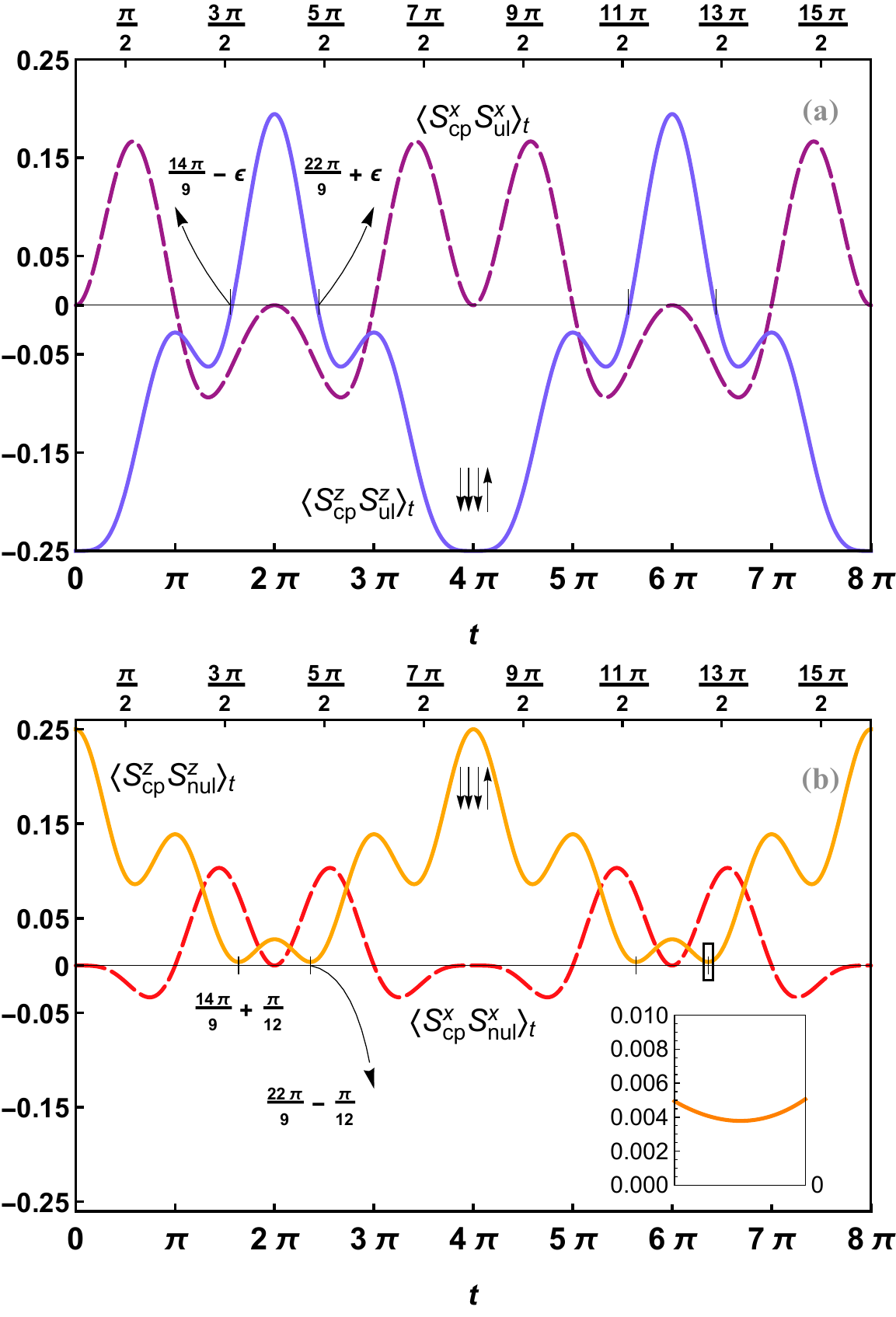}
		\caption{Correlation functions of different spin directions between (a) CP and UL (with $\epsilon=\frac{17}{1000}\pi$), (b) CP and a NUL.}
		\label{fig:corulcpnul}
	\end{minipage}
	\hfill
	\begin{minipage}{0.49\textwidth}
		\includegraphics[width=\textwidth]{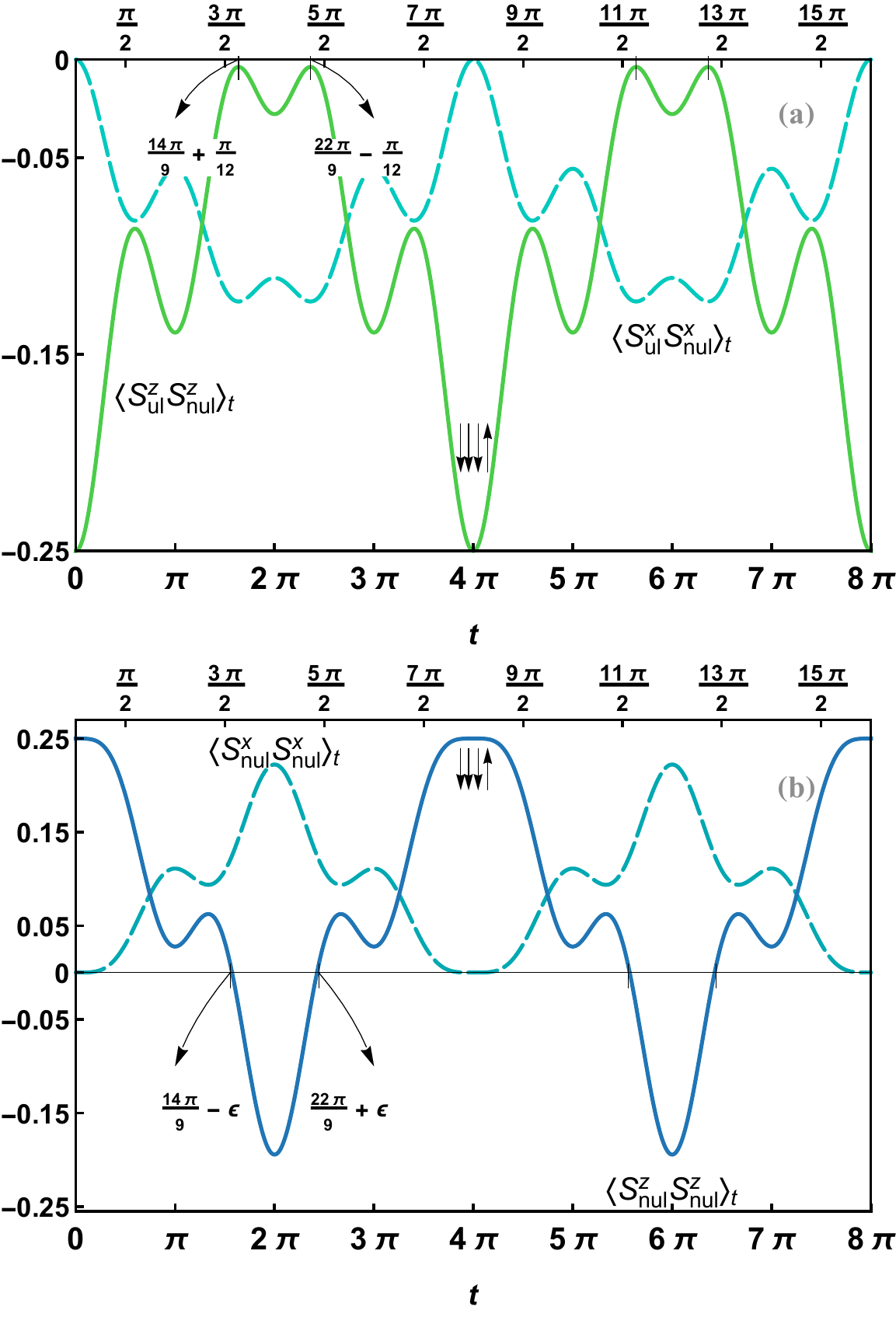}
		\caption{Correlation functions of different spin directions between (a) the UL and a NUL, (b) two NULs (with $\epsilon=\frac{17}{1000}\pi$).
		}
		\label{fig:corulnulul}
	\end{minipage}
\end{figure}

Lastly, we address to calculate two-point quantum correlations ($
W^{\alpha\alpha}_{\imath\jmath}(t)=\langle S_{\imath}^{\alpha}S_{\jmath}^{\alpha}\rangle_{t}$). Two-point correlations between any two qubits in the $z$-direction are given by Eqs.~\ref{eq:0001tcor2tayi}, 

\begin{eqnarray}
\langle S^{z}_{cp}S^{z}_{ul}\rangle_t&=&\frac{1}{72}\Big[{-12}\cos \left(\frac{t}{2}\right)-4 \cos \left(\frac{3 t}{2}\right)
+3\cos (2 t)-5\Big]\notag\\
\langle S^{z}_{cp}S^{z}_{nul}\rangle_t&=&\frac{1}{72}\Big[{6}\cos \left(\frac{t}{2}\right)+2 \cos \left(\frac{3 t}{2}\right)
+3\cos (2 t)+7\Big]\notag\\
\langle S^{z}_{ul}S^{z}_{nul}\rangle_t&=&-\langle S^{z}_{cp}S^{z}_{nul}\rangle_t\notag\\
\langle S^{z}_{nul}S^{z}_{nul}\rangle_t&=&-\langle S^{z}_{cp}S^{z}_{ul}\rangle_t,
\label{eq:0001tcor2tayi}
\end{eqnarray}  
and in the $x$-direction are written in Eqs.~\ref{eq:0001tcor2tayix}.
\begin{eqnarray}
\langle S^{x}_{cp}S^{x}_{ul}\rangle_t&=&\frac{1}{12}\Big[\sin ^2\left(\frac{t}{2}\right) \left(4 \cos \left(\frac{t}{2}\right)+\cos (t)+1\right)\Big] \notag\\
\langle S^{x}_{cp}S^{x}_{nul}\rangle_t&=&-\frac{1}{6}\Big[\sin ^2\left(\frac{t}{4}\right) \sin \left(\frac{t}{2}\right) \sin (t)\Big] \notag\\
\langle S^{x}_{ul}S^{x}_{nul}\rangle_t&=&\frac{1}{144}\Big[{6}\cos \left(\frac{t}{2}\right)+2 \cos \left(\frac{3 t}{2}\right) 
+3\cos (2 t)-11\Big]\notag\\
\langle S^{x}_{nul}S^{x}_{nul}\rangle_t&=&\frac{1}{9}\Big[\sin ^4\left(\frac{t}{2}\right) \left(8 \cos \left(\frac{t}{2}\right)+3\cos (t)+7\right)\Big].
\label{eq:0001tcor2tayix}
\end{eqnarray}  

Figs.~\ref{fig:corulcpnul} and Figs.~\ref{fig:corulnulul} respectively present Eqs.~\ref{eq:0001tcor2tayi} and~\ref{eq:0001tcor2tayix}. 
Considering to Figs.\ref{fig:corulcpnul}~(a) and \ref{fig:corulnulul}~(b) we find that there is a mirror symmetry for two-point correlations in the $z$-direction for CP-UL and NUL-NUL pairs. In the opposite to $z$-direction correlations in the COPS time evolution, these correlations are not zero at $\tau^{**}$. Actually, they are zero at $\tau^{**}_n\pm \epsilon$ (with $\epsilon=\frac{17}{1000}\pi$). 

Moreover, as shown in Figs.~\ref{fig:corulcpnul}~(b) and~\ref{fig:corulnulul}~(a), 
$\langle S^{z}_{ul}S^{z}_{nul}\rangle_t=-\langle S^{z}_{cp}S^{z}_{nul}\rangle_t$ for CP-NUL and UL-NUL pairs. These functions are close to zero at $\tau^{**}_n\pm \frac{\pi}{12}$ (see the inset of Fig.~\ref{fig:corulcpnul}~(b)), but like two other pairs are not zero exactly at PSTWS.

\section{conclusion}
\label{sec:conclu}

In the present work, we have studied the time dependent pairwise entanglement and two-point quantum correlations in the XXX-3LSSS for two different scenarios. At the first stage, the COPS was applied upon by the time evolution operator. An operation ends up having a quantum state which is extended on all system's one-particle states.
We found that $W$ states can be generated at TWS (which are given by $t=\frac{(2n-1)\pi}{2}$). It is shown that all one-particle sates can be present with the same probabilities at TWS. Additionally, the expectation value of the local $z$-spin component ($\langle S_{i}^{z}\rangle_{t})$ of the different qubits are all equal at TWS. It is found that the equal entanglement among different qubits at TWS is raised from $xx$ and $yy$ two-point correlation functions because the $zz$ two point correlation vanishes at these time instants. In this scenario, all elements of the structure disentangle from each other at $t=n\pi$.

In the second scenario, an arbitrary LOPS was applied upon by the time evolution operator. In this situation, $W$ states don't have any chances to be generated, although all concurrences meet closely each other at PSTWS which are given by $t^{**}_{n}=\pi( 2 n-\frac{(9+5(-1)^{n})}{9})$. In addition, the probabilities of different one-particle states and the expectation value of local $z$-spin components ($\langle S_{i}^{z}\rangle_{t})$s are not equal to each other. Moreover, the $zz$ quantum correlation between any two qubits does not vanish at PSTWS. 

It should be pointed out that independent of whether the initial state is the COPS or a LOPS, the central qubit disentangles from the ligands at the same time instants, however disentangling between all elements strongly depends on the selected initial one-particle state.

\section*{Acknowledgments}
This work is based upon research supported by the
Shahid Bahonar University Foundation. 
The author would like to thank Dr. Baghani, Dr. Davody, F. Mohtashamian, Dr. Maleki for their assistance in reviewing this manuscript. 
The author thanks Changhua Zhu and other scientists who helped and communicated for clarifying some details and approaches in the past years specially Saeed Mahdavifar, Andreas Honecker and Mathias Troyer.


\end{document}